\documentclass[a4paper]{article}

\usepackage{19lomcon}        % Proceedings volume layout metrics
\usepackage{cite}             % Smart range citations
\usepackage{epsfig}           % Encapsulated PostScript figure inclusion
\usepackage{epstopdf}
\usepackage{bm}

\begin{document}

%%%%%%%%%%%%%%%%%%%%%%%%%%%%%%%%%%%%%%%%%%%%%%%%%%%%%%%%%%%%%%%%%%
% The preamble of the paper
%%%%%%%%%%%%%%%%%%%%%%%%%%%%%%%%%%%%%%%%%%%%%%%%%%%%%%%%%%%%%%%%%%

\title{THE THREE-LOOP ADLER $D$-FUNCTION FOR ${\cal N}=1$
SQCD WITH VARIOUS RENORMALIZATION PRESCRIPTIONS}

\author{S.S.Aleshin, \email{aless2001@mail.ru}}
\affiliation{
{\small{\em Institute for Information Transmission Problems RAS}},\\
{\small{\em Bolshoy Karetny per. 19, Moscow, 127051, Russia}},\\}
\author{A.L.Kataev, \email{kataev@m2.inr.ac.ru}}
\affiliation{
{\small{\em Institute for Nuclear Research RAS,}}\\
{\small{\em 60th October Anniversary Prospect 7a, Moscow 117312, Russia}};\\
{\small{\em Moscow Institute of Physics and Technology,}}\\
{\small{\em Institutskiy per. 9. Dolgoprudny. Moscow Region. 141700,   Russia}},\\}
\author{K.V.Stepanyantz \email{stepan@m9com.ru}}
\affiliation{
{\small{\em Moscow State University, Faculty of Physics}}\\
{\small{\em Leninskie Gory, Moscow, 119991,  Russia}}}

% You may repeat \author and \affiliation as many times as necessary!

\date{}
% Print it out!
\maketitle

%%%%%%%%%%%%%%%%%%%%%%%%%%%%%%%%%%%%%%%%%%%%%%%%%%%%%%%%%%%%%%%%%%
% The preamble of the paper
%%%%%%%%%%%%%%%%%%%%%%%%%%%%%%%%%%%%%%%%%%%%%%%%%%%%%%%%%%%%%%%%%%

\begin{abstract}
The three-loop Adler $D$-function for ${\cal N}=1$ SQCD in the $\overline{\mbox{DR}}$ scheme is calculated. It appears that the result does not satisfy NSVZ-like equation which relates the $D$-function to the anomalous dimension of the matter superfields. However this NSVZ-like equation can be restored by a special tuning of the renormalization scheme. Also we demonstrate that the $D$-function defined in terms of the bare coupling does not satisfy the NSVZ-like equation in the case of using the regularization by dimensional reduction. The scheme-dependence of the $D$-function written in the form of the $\beta$-expansion is briefly discussed.
\end{abstract}

\section{Introduction}
The $D$-function allows comparing the theoretical QCD predictions with the experimental data for $R$-ratio which plays an important role in investigating strong interaction contributions to various physical quantities (such as the strong coupling constant or the muon anomalous magnetic moment).

In the region where the perturbation theory is applicable the $D$-function can be found by calculating the QCD corrections to the electromagnetic coupling constant \cite{Adler:1974gd}
\begin{equation}\label{D_Function_O}
\bm{D}(\alpha_s) = - \frac{3\pi}{2}\frac{\partial}{\partial\ln P} d^{-1}(\alpha,\alpha_{s},P/\mu)
\Big|_{\alpha\to 0},
\end{equation}
where $d^{-1}$ is the inverse invariant charge. In this equation the electromagnetic coupling constant is set to zero. Therefore, only quantum corrections coming from the quark and gluon loops are taken into account, while the electromagnetic field is treated as an external one, so that only QCD corrections survive. Sometimes, it is also convenient to use another definition of the $D$-function,
\begin{equation}\label{D_Function_R}
\widetilde D(\alpha_s) = -\frac{3\pi}{2}\frac{d}{d\ln\mu}\alpha^{-1}(\alpha_0,\alpha_{s0},\Lambda/\mu)
\Big|_{\alpha_0,\alpha_{s0} = \mbox{\scriptsize const};\ \alpha_0\to 0}.
\end{equation}
It is possible to demonstrate \cite{Aleshin:2019yqj} that $\bm{D}(\alpha_s) = \widetilde D(\alpha_s)$ if the SQCD-renormalization of the electromagnetic coupling constant is made according to the prescription
\begin{equation}\label{MOM_Like_Scheme}
\Pi(\alpha_s, P/\mu=1) = 0.
\end{equation}

According to \cite{Shifman:2014cya,Shifman:2015doa} the $D$-function defined in terms of the bare coupling constant in all orders satisfies the NSVZ-like equation
\begin{equation}\label{NSVZ_Like_Relation_Bare}
D(\alpha_{0s}) = \frac{3}{2} \sum_{\alpha=1}^{N_f}q_{\alpha}^2\, \Big(\mbox{dim}(R) - \mbox{tr}\, \widetilde\gamma(\alpha_{0s})\Big)
\end{equation}
if the higher covariant derivatives \cite{Slavnov:1971aw,Slavnov:1972sq} are used for the regularization independently of a renormalization prescription supplementing it. For the $D$-function (\ref{D_Function_R}) a similar exact relation takes place in the HD+MSL scheme, when only powers of $\ln\Lambda/\mu$ are included into renormalization constants \cite{Kataev:2017qvk}.

However, the prescription (\ref{MOM_Like_Scheme}) giving the Adler function (\ref{D_Function_O}) evidently differs from the HD+MSL scheme. Moreover, most calculations of the phenomenological interest have been done in the $\overline{\mbox{DR}}$-scheme, so that it is desirable to find the result for the $D$-functions (\ref{D_Function_O}) and (\ref{D_Function_R}) in this case as well.

\section{The 3-loop results for Adler functions $\bm{D}(\alpha_s)$ and $\widetilde D(\alpha_s)$}
We consider massless ${\cal N}=1$ SQCD with the $G\times U(1)$ gauge group. This theory contains two bare coupling constants $g_0$ and $e_0$ corresponding to the factors $G$ and $U(1)$, respectively. We formulate it in terms of the ${\cal N}=1$ superfields,
\begin{eqnarray}\label{Action_Classical_SQCD}
&& S = \frac{1}{2 g_0^2}\,\mbox{Re}\,\mbox{tr}\int d^4x\, d^2\theta\,W^a W_a
+ \frac{1}{4 e_0^2}\,\mbox{Re}\int d^4x\, d^2\theta\,\bm{W}^a \bm{W}_a \qquad\nonumber\\
&& +\smash{\sum_{\alpha=1}^{N_{f}}} \frac{1}{4}\,\int d^4x\, d^4\theta\,\Big(\phi_{\alpha}^{+}e^{2V+2q_{\alpha}\bm{V}}\phi_{\alpha}
+ \widetilde\phi_{\alpha}^{+}e^{-2V^t-2q_{\alpha}\bm{V}}\widetilde\phi_{\alpha}\Big),
\end{eqnarray}
where $\phi_{\alpha}$, $\widetilde{\phi}_{\alpha}$ are chiral matter superfields  in the representations $R$ and $\bar{R}$, respectively. $V$ is the non-Abelian gauge superfield, and $\bm{V}$ is the Abelian one. The corresponding gauge superfield strengths are denoted by $W_a$ and $\bm{W}_a$.

The result for the three-loop $D$-function obtained in \cite{Kataev:2017qvk} with the help of the higher covariant derivative regularization can be written as
\begin{eqnarray}\label{D_Function_Higher_Derivatives}
&&\hspace*{-9mm} \widetilde D(\alpha_s)=
\frac{3}{2}\smash{\sum_{\alpha=1}^{N_{f}}}q_{\alpha}^2\,\Big\{\mbox{dim}(R)+\frac{\alpha_s}{\pi}\mbox{tr}\,C(R) + \frac{\alpha^2_s}{\pi^2}
\Big[\,\frac{3}{2}C_2\mbox{tr}\,C(R)\Big(\ln a_{\varphi}+1+d_2 \nonumber\\
&&\hspace*{-9mm} -b_{11}\Big) - N_{f}T(R)\,\mbox{tr}\, C(R)\Big(\ln a+1+d_2-b_{12}\Big)
-\frac{1}{2}\mbox{tr}\,\big(C(R)^2\big)\Big]\Big\} + O(\alpha_s^3),
\end{eqnarray}
where $r = \dim G$ and
\begin{equation}
\mbox{tr}\,(T^{A}T^{B}) = T(R)  \delta^{AB};\quad\ C(R)_i{}^j = (T^AT^A)_i{}^j;\quad C_2\delta^{CD} = f^{ABC}f^{ABD}.
\end{equation}
In Eq. (\ref{D_Function_Higher_Derivatives}) $a = M/\Lambda$ and $a_\varphi = M_\varphi/\Lambda$ are the ratios of the Pauli--Villars masses (needed for regularizing the one-loop divergences \cite{Slavnov:1977zf}) to the dimensionful regularization parameter. The finite constants $b_{11}$, $b_{12}$, and $d_2$ characterize the subtraction scheme. Fixing the renormalization prescription we fix values of these constants defined by the equations
\begin{eqnarray}
&&\hspace*{-4mm} \frac{1}{\alpha_{0s}}= \frac{1}{\alpha_{s}} + \frac{1}{2\pi}\ \Big[3C_2\Big(\ln\frac{\Lambda}{\mu}+b_{11}\Big)-2N_fT(R)\Big(\frac{\Lambda}{\mu}+b_{12}\Big)\Big] + O(\alpha_s);\nonumber\\
&&\hspace*{-4mm} \frac{1}{\alpha_{0}} = \frac{1}{\alpha} - \frac{1}{\pi}\sum_{\alpha=1}^{N_{f}}q_{\alpha}^2\, \mbox{dim}(R)
\Big(\ln\frac{\Lambda}{\mu}+d_1\Big) - \frac{\alpha_s}{\pi^2}\sum_{\alpha=1}^{N_{f}}q_{\alpha}^2\, \mbox{tr}\,C(R)\Big(\ln\frac{\Lambda}{\mu}+d_2\Big)\nonumber\\
&& \hspace*{-4mm} +O(\alpha_s^2).
\end{eqnarray}
The results for the functions (\ref{D_Function_O}) and (\ref{D_Function_R}) with the dimensional reduction \cite{Siegel:1979wq} are obtained from Eq. (\ref{D_Function_Higher_Derivatives}) for some values of these finite constants. These values can be found by comparing the expressions for the two-loop anomalous dimensions of the matter superfields and also the two-point Green functions of the matter superfields and of the electromagnetic gauge superfield calculated with the higher derivative regularization and with a proper renormalization prescription supplementing dimensional reduction, see \cite{Aleshin:2019yqj} for details.

It is convenient to present the resulting expressions in the form of the $\beta$-expansion formalism \cite{Kataev:2016aib}. It is also reasonable to use a similar expansion for the anomalous dimension of the matter superfields in order that both sides of the NSVZ-like relation have a similar structure,
\begin{eqnarray}
&&D(\alpha_s) = \frac{3}{2}\sum\limits_{\alpha=1}^{N_f} q_{\alpha}^2\, \Big\{\ \mbox{dim}(R)+ \mbox{tr}\, C(R) \sum\limits_{n=1}^{2} D_n \Big(\frac{\alpha_s}{\pi}\Big)^n \ \Big\} + O(\alpha_s^3);\nonumber\\
&& \widetilde\gamma(\alpha_s)_i{}^j = \sum\limits_{n=1}^{2} \big(\gamma_n\big)_i{}^j \Big(\frac{\alpha_s}{\pi}\Big)^n + O(\alpha_s^3),
\end{eqnarray}
where $\beta_0=-3C_2/2+N_fT(R)$ is the first coefficient of the $\beta$-function and
\begin{eqnarray}
&& D_1 = D_1[0],\qquad\qquad\quad D_2 = \beta_0 D_2[1] + D_2[0]\;\qquad\nonumber\\
&& \big(\gamma_1\big)_i{}^j = \gamma_1[0]_i{}^j,\qquad\,\quad \big(\gamma_2\big)_i{}^j = \beta_0\, C(R)_i{}^j\, \gamma_2[1] + \gamma_2[0]_i{}^j.\qquad
\end{eqnarray}
The scheme-independent coefficient are
\begin{eqnarray}
&&\hspace*{-4mm} D_1[0] = 1;\qquad\qquad\quad\quad \ D_2[0] = -\frac{\mbox{tr}(C(R)^2)}{2\, \mbox{tr}\,C(R)};\nonumber\\
&&\hspace*{-4mm} \gamma_1[0]_i{}^j = - C(R)_i{}^j;\qquad\quad \gamma_2[0]_i{}^j = \frac{1}{2} \left(C(R)^2\right)_i{}^j,
\end{eqnarray}
while the values of the scheme-dependent coefficients $D_2[1]$ and $\gamma_2[1]$ are collected in Table 1 (see \cite{Aleshin:2019yqj} for details). From this table we see that NSVZ-like equation for the $D$-function is not satisfied in the $\overline{\mbox{DR}}$-scheme. However, it is possible to construct finite renormalizations that restore it. The NSVZ-like relation is not satisfied for the $D$-function defined in terms of the bare coupling constant for the theory  regularized by dimensional reduction, in contrast to the case of higher covariant derivatives for which it is satisfied.
\begin{table}[h]
\tiny\
\begin{tabular}{|c|c|c|c|c|c|c|}
\hline
Function & Regularization & $\alpha$ & $\alpha_s$ and $Z_i{}^j$ & $D_2[1]\vphantom{\Big(}$ & $\gamma_2[1]$ & NSVZ\\
\hline\hline
$\vphantom{\Big(} \bm{D}(\alpha_s)$ & DRED & MOM-like & $\overline{\mbox{DR}}$ & ${\displaystyle -\frac{5}{2}+ \frac{3}{2}\zeta(3)}\vphantom{\Big(}$ & ${\displaystyle \frac{1}{2}\vphantom{\Big(}}$ & $-$\\
\hline
$\vphantom{\Big(} \bm{D}(\alpha_s)$ & DRED & MOM-like & $\mbox{NSVZ}$ & ${\displaystyle -\frac{5}{2}+\frac{3}{2}\zeta(3) + \widetilde\delta_0\vphantom{\Big(}}$ & ${\displaystyle \, \frac{5}{2} - \frac{3}{2}\zeta(3) - \widetilde\delta_0\vphantom{\Big(}}$ & $+$\\
\hline
$\vphantom{\Big(} \widetilde{D}(\alpha_s)$ & DRED & $\overline{\mbox{DR}}$ & $\overline{\mbox{DR}}$ & ${\displaystyle -\frac{3}{4}\vphantom{\Big(}}$ & ${\displaystyle \frac{1}{2}\vphantom{\Big(}}$ & $-$\\
\hline
$\vphantom{\Big(} \widetilde{D}(\alpha_s)$ & DRED & \parbox{1.5cm}{$\overline{\mbox{DR}}$+finite renormal.} & $\mbox{NSVZ}$ & ${\displaystyle  -\frac{3}{4} + \widetilde\delta_0 - \widetilde f_1}\vphantom{\Big(}$ & ${\displaystyle \frac{3}{4} - \widetilde\delta_0 + \widetilde f_1}$ & $+$\\
\hline
$\vphantom{\Big(} \widetilde{D}(\alpha_s)$ & HD; $a_\varphi = a$ & HD+MSL & HD+MSL & ${\displaystyle -1-\ln a\vphantom{\Big(}}$ & ${\displaystyle 1+\ln a\vphantom{\Big(}}$ & $+$\\
\hline
$\vphantom{\Big(} D(\alpha_{s0})$ & DRED & arbitrary & arbitrary & ${\displaystyle -\frac{3}{4} -\frac{1}{2\varepsilon}}$ & ${\displaystyle \frac{1}{2}\vphantom{\Big(}}$ & $-$\\
\hline
$\vphantom{\Big(} D(\alpha_{s0})$ & HD; $a_\varphi = a$ & arbitrary & arbitrary & ${\displaystyle -1-\ln a\vphantom{\Big(}}$ & ${\displaystyle 1+\ln a\vphantom{\Big(}}$ & $+$\\
\hline
\end{tabular}
\normalsize
\caption{Scheme-dependent coefficients of the $\beta$-expansions of the $D$-function and of the anomalous dimension of the matter superfields in various renormalization schemes.}\label{Table_Scheme_Dependence}
\end{table}

The authors are very grateful to A.E.Kazantsev for valuable discussions.
The work of A.K. and K.S. was supported by the Foundation for the Advancement of Theoretical Physics and Mathematics `BASIS', grants No. 17-11-120-1 (A.K.) and 19-1-1-45-1 (K.S.).

%%%%%%%%%%%%%%%%%%%%%%%%%%%%%%%%%%%%%%%%%%%%%%%%%%%%%%%%%%%%%%%%%%
% References
%%%%%%%%%%%%%%%%%%%%%%%%%%%%%%%%%%%%%%%%%%%%%%%%%%%%%%%%%%%%%%%%%%

\end{document}